\documentclass[usenatbib]{mn2e}

\usepackage{amsmath}  
\usepackage{amssymb}  
\usepackage{mathrsfs}
\usepackage[all]{xy}
\usepackage[pdftex]{graphicx}
\usepackage{indentfirst}
\usepackage[pdftex]{hyperref}
\usepackage{verbatim}
\usepackage{float}
\usepackage{subfig}

\newcommand{\Ts}{T_{\text{S}}}
\newcommand{\Tk}{T_{\text{K}}}
\newcommand{\Tb}{T_{\text{b}}}
\newcommand{\Tx}{T_{\text{X}}}
\newcommand{\Tvir}{T_{\text{vir}}}

\newcommand{\OL}{\Omega_{\Lambda}}
\newcommand{\Om}{\Omega_{\text{m}}}
\newcommand{\Ob}{\Omega_{\text{b}}}
\newcommand{\Ok}{\Omega_{\text{k}}}
\newcommand{\Ox}{\Omega_{\text{X}}}

\newcommand{\MJ}{M_{\text{J}}}
\newcommand{\Msf}{M_{\text{sf}}}
\newcommand{\Mmin}{M_{\text{min}}}
\newcommand{\Msun}{\text{M}_{\odot}}

\newcommand{\Deltac}{\Delta_{\text{c}}}
\newcommand{\ns}{n_{\text{s}}}
\newcommand{\mx}{m_{\text{X}}}
\newcommand{\gx}{g_{\text{X}}}
\newcommand{\Rc}{R_{\text{c}}^0}

\newcommand{\zh}{z_{\text{h}}}
\newcommand{\zmin}{z_{\text{min}}}
\newcommand{\zr}{z_{\text{r}}}

\newcommand{\cmfast}{\textsc{\small 21CMFAST}}

\begin{document}

\title[Imprint of Warm Dark Matter on 21-cm Signal]{The Imprint of Warm Dark Matter on the Cosmological 21-cm Signal}

\author[Sitwell et al.]{Michael Sitwell$^1$, Andrei Mesinger$^2$, Yin-Zhe Ma$^1$, \& Kris Sigurdson$^1$\\
$^1$Department of Physics and Astronomy, University of British Columbia, Vancouver, BC, V6T 1Z1, Canada\\
$^2$Scuola Normale Superiore, Piazza dei Cavalieri 7, 56126 Pisa, Italy}

\voffset=-0.6in

\maketitle

\begin{abstract}

We investigate the effects of warm dark matter (WDM) on the cosmic 21-cm signal. If dark matter exists as WDM instead of cold dark matter (CDM), its non-negligible velocities can inhibit the formation of low-mass halos that normally form first in CDM models, therefore delaying star-formation. The absence of early sources delays the build-up of UV and X-ray backgrounds that affect the 21-cm radiation signal produced by neutral hydrogen. With use of the \cmfast \, code, we demonstrate that the pre-reionization 21-cm signal can be changed significantly in WDM models with a free-streaming length equivalent to that of a thermal relic with mass $\mx$ of up to $\sim 10$--$20$ keV. In such a WDM cosmology, the 21-cm signal traces the growth of more massive halos, resulting in a delay of the 21-cm absorption signature and followed by accelerated X-ray heating. CDM models where astrophysical sources have a suppressed photon-production efficiency can delay the 21-cm signal as well, although its subsequent evolution is not as rapid as compared to WDM. This motivates using the gradient of the global 21-cm signal to differentiate between some CDM and WDM models. Finally, we show that the degeneracy between the astrophysics and $\mx$ can be broken with the 21-cm power spectrum, as WDM models should have a bias-induced excess of power on large scales. This boost in power should be detectable with current interferometers for models with $\mx\lesssim 3\,\text{keV}$, while next generation instruments will easily be able to measure this difference for all relevant WDM models.

\end{abstract}


\section{Introduction}

Hierarchical structure formation within the $\Lambda$CDM model has been exceptionally accurate in describing the large-scale Universe within the range $\sim 10 \,\text{Mpc} - 1 \,\text{Gpc}$, as demonstrated from studies of the cosmic microwave background (CMB) and the clustering of galaxies. However, for over a decade concerns have been raised over whether the standard assumption of cold dark matter (CDM) provides an adequate fit to data on smaller, sub-Mpc scales. These include predictions from $N$-body simulations that yield an overabundance of galactic satellites in our galaxy and in the field (\citealt{moore1999, klypin1999, Papastergis2011}), as well as in voids (\citealt{peebles2001}), and produce overly-dense galactic centres with `cuspy' density profiles (\citealt{deblok2001, donato2009, newman2009}) and are inconsistent with observations of the kinetic properties of bright Milky Way satellites (\citealt{boylan2011, boylan2012}). 

One possible explanation lies with baryonic feedback processes (\citealt{governato2007, pontzen2012, kimmel2013, sobacchi2013, teyssier2013}), although accurately modelling these mechanisms is often challenging and difficulties may persist in matching to observations.

Another possible explanation is to change the properties of dark matter so it is warm (WDM).\footnote{Other possible alterations to the standard CDM model that may resolve these small-scale problems include self-interacting dark matter (\citealt{spergel2000,burkert2000,dave2001}) and atomic dark matter or other models with acoustic damping of dark matter fluctuations (\citealt{kaplan2010,cyrracine2013}).} This may alleviate these small-scale problems due to the higher velocities of the dark matter. In this case, structures are smoothed on scales below the dark matter's free-streaming length. Non-relativistic residual velocities can delay halo collapse and star formation.  These effects may reduce the number of sub-haloes and low-mass galaxies that are formed as well as flatten out galactic centres.

The two most popular WDM candidates in the literature motivated by particle physics have been the sterile neutrino (\citealt{dodelson1994, abazajian2001, boyarsky2009}) and the gravitino (\citealt{bond1982, pagels1982}). While WDM may be produced in a number of different ways, it is most often described as a thermal relic that decouples while relativistic, but is non-relativistic by matter-radiation equality as to preserve structure beyond the Mpc scale. In this case, the WDM would have a particle mass $\mx$ of the order of a keV. Although for our purposes the free-streaming scale of the dark matter is a more fundamental quantity, we use the standard convention of discussing the WDM mass of a thermal relic instead. We caution that for other WDM production mechanisms the correspondence between free-streaming length and mass will be different. We also remark that the results presented in this paper can be applicable to models other than WDM that have similar cut-off scales in their power spectrum (see, e.g. \citealt{cyrracine2013}).

As WDM suppresses growth of small structures, which form first in the hierarchical structure formation of CDM, early star formation is delayed in WDM models. Detection of signals emitted from high-redshift objects either directly, such as from gamma-ray bursts (GRBs) (\citealt{mesinger2005}) or strongly lensed galaxies (\citealt{pacucci2013}), or indirectly through the redshift of reionization (\citealt{barkana2001}), can place constraints on $\mx$. Recently, \citealt{desouza2013} using GRB catalogues placed a constraint of $\mx>1.6-1.8 \,\text{keV}$ at 95\% CL. Requiring WDM models to be able to reproduce both the stellar mass function and Tully-Fisher relation places a lower bound of $\mx\geq 0.75 \,\text{keV}$ (\citealt{kang2013}). The Lyman-$\alpha$ forest can probe scales down to $\sim 1 \,\text{Mpc}$ and can provide strict limits on $\mx$ (\citealt{narayanan2000, seljak2006, viel2005, viel2008}), with the most recent and stringent constraint of $\mx>3.3 \,\text{keV}$ at $2\sigma$ (\citealt{viel2013}). Although it has been claimed that the less dense galactic cores formed in WDM models may provide a better fit to the kinematic data of bright Milky Way satellites (\citealt{lovell2012}), there is an ongoing debate as to whether WDM with a mass above current lower bounds can create a large enough galactic core as needed to solve the `cusp-core' problem (\citealt{navarro2011, maccio2012}; though see \citealt{devega2013}).

Highly-redshifted 21-cm radiation emitted from the hyperfine spin-flip of neutral hydrogen is a promising new tool to probe the high-redshift Universe (\citealt{madau1997,furlanetto2006,zaldarriaga2004,morales2010,mesinger2013a}). If WDM is present in sufficient quantities to significantly delay structure formation, it could potentially leave a trace within the 21-cm radiation signal. Light emitted by the first astrophysical sources can couple the spin temperature of neutral hydrogen to the kinetic temperature of the IGM through the Wouthuysen-Field (WF) mechanism (\citealt{wouthuysen1952,field1958}), as well as heat and ionize the IGM. Thus, a delay in the appearance of these early sources can alter the 21-cm signal and delay milestones in the signal. In this paper, we will examine the effects of WDM on the pre-reionization 21-cm signal. This era may be especially useful for examining WDM since WDM inhibits the formation of low-mass halos that form first in CDM models and thus differences between the halo populations in CDM and WDM increase with redshift. As astrophysics is very poorly known at high-redshifts ($z\geq 6$), \textit{we will focus on characterizing degeneracies between the unknown astrophysics and the presence of WDM}.

The outline of this paper is as follows: In Section~\ref{wdm}, we review the effects of the free-streaming of the WDM on the linear power spectrum and its residual velocities on halo collapse. The basic properties of the 21-cm signal are outlined in Section~\ref{21cmsignal} and its simulation is described in Section~\ref{21cmmodel}, with the simulation results discussed in Section~\ref{results}. Throughout this paper, we assume cosmological parameter values of $\OL=0.73,\, \Om=0.27,\, \Ob=0.046,\, h=0.7,\, \sigma_8=0.82,\, \ns=0.96$. We quote all quantities in comoving units, unless stated otherwise.

\section{Effect of WDM on structure formation}
\label{wdm}

\subsection{Free-streaming}

The free-streaming of WDM particles smears out perturbations on small scales, as WDM particles stream out of over-dense regions and into under-dense regions. Perturbations are suppressed on scales below that corresponding to the WDM particle horizon.

The effect of free-streaming on the spectrum of linear perturbations can be included by use of a transfer function $\Tx(k)$ that dampens small-scale fluctuations as compared to those in CDM. This transfer function can be found by fitting the results of a Boltzmann code, which we take as
\begin{equation}
\Tx(k) = (1 + (\epsilon k \Rc)^{2\nu} )^{-\eta/\nu}
\label{TX}
\end{equation}
where $\epsilon=0.361$, $\eta=5$, and $\nu=1.2$ (\citealt{bode2001}, hereafter BHO). $\Rc$ is the comoving cutoff scale, at which the power in $k=1/\Rc$ is reduced by half compared to that in CDM, and is given by
\begin{equation}
\Rc = 0.201 \left(\frac{\Ox h^2}{0.15}\right)^{0.15} \left(\frac{\gx}{1.5}\right)^{-0.29} \left(\frac{\mx}{\text{keV}}\right)^{-1.15} \text{Mpc}
\end{equation}
where $\gx$ is the number of effective degrees of freedom contributing to number density, with bosons contributing unity to $\gx$ and fermions contributing $3/4$. We will use the standard assumption that the WDM is a spin-$\frac{1}{2}$ fermion, so that $\gx=3/2$. $\Ox$ is the energy density parameter contributed by the WDM, which we set to $\Ox=\Om-\Ob$ as we will only be considering models where WDM constitutes the whole of the dark matter. The transfer function in Eq.~(\ref{TX}) serves to suppress small-scale linear perturbations in the power spectrum, which we generate using the transfer function of \citealt{eisenstein1998}.

 \subsection{Residual velocities}
 \label{resvel}
 
In addition, the residual velocity dispersion of the WDM delays the growth of non-linear perturbations and consequently collapse into virialized halos. This can be thought of as an `effective pressure'. BHO modelled the collapse in WDM by studying collapse in an analogous system comprised of an adiabatic gas, so its root-mean-square velocity evolves as $v_{\text{rms}}\propto 1/a$, as the case with WDM, and whose initial temperature is set such that it shares the same  $v_{\text{rms}}$ with the WDM.

Using the gas analogue in a spherically symmetric hydrodynamics simulation, BHO computed the linear collapse threshold $\delta_c(M,z)$, finding that the collapse threshold rises sharply near the Jeans mass $\MJ$ for the analogue gas. The results of \citealt{desouza2013} showed that using the extended Press-Schechter (EPS) formalism to compute the collapse fraction with a sharp minimum mass cutoff at $\MJ$ and the collapse threshold for spherical collapse in CDM ($\delta_c\approx1.69$) is in good agreement with the full random-walk procedure with the WDM modified collapse threshold as used in BHO. To achieve this close agreement, a factor of 60 was added to the expression for $\MJ$ originally found in BHO, so that $\MJ$ is given by
\begin{equation}
\MJ \approx 1.5\times 10^{10} \left(\frac{\Ox h^2}{0.15}\right)^{1/2} \left(\frac{\mx}{\text{keV}}\right)^{-4} \Msun
\label{MJwdm}
\end{equation}
As using the sharp cutoff at $\MJ$ is much less computationally intensive and easily integrable within the EPS formalism, we employ this method instead of the full random-walk procedure.

\subsection{Halo Abundances}
\label{halos}

The production rate of photons that are capable of heating or ionizing the IGM, or coupling the spin temperature to the colour temperature via the WF mechanism, is modelled as being proportional to the collapse fraction $f_{\text{coll}}(z, \Mmin)$ of halos with sufficient mass ($\geq \Mmin$) to host star-forming galaxies. To compute the mean collapse fraction, we use the Sheth-Tormen mass function (\citealt{sheth2001}), giving the comoving number density of halos with mass between $M$ and $M+dM$ as
\begin{equation}
\frac{dn_{\text{ST}}}{dM} = - A \sqrt{\frac{2}{\pi}} \frac{\bar{\rho}_{\text{m}}}{M} \frac{d\text{ln}\sigma}{dM} \hat{\nu} (1 + \hat{\nu}^{-2p}) e^{-\hat{\nu}^2/2}
\end{equation}
where $\hat{\nu}=\sqrt{a}\delta_c(M,z)/\sigma(M)$, $\bar{\rho}_{\text{m}}$ is the mean matter energy density, $\sigma(M)$ is the rms of density fluctuations smoothed on a scale that encompasses a mass $M$. $A$, $a$, and $p$ are fit parameters taken as $A=0.353$, $a=0.73$, and $p=0.175$ (\citealt{jenkins2001}). The mean collapse fraction is computed as
\begin{equation}
f_{\text{coll}}(>\Mmin,z) = \frac{1}{\rho_{\text{m}}} \int_{\Mmin}^{\infty} M \frac{dn_{\text{ST}}}{dM} dM
\label{fcoll}
\end{equation}
where $\Mmin = \text{max}(\MJ, \Msf)$ and $\Msf$ is the minimum halo mass where star-formation can occur. $\MJ$ is assigned a value of zero in the case of CDM. It will be convenient to express $\Msf$ in terms of the corresponding virialized halo temperature $\Tvir$ as (see, for instance,  \citealt{barkanaloeb2001})
\begin{multline}
\Msf = 9.37 \times 10^7  \left(\frac{\mu}{0.6}\right)^{-3/2} \left(\frac{h}{0.7}\right)^{-1} \left(\frac{\Om}{0.3}\right)^{-1/2} \\ \times  \left(\frac{1}{\Om^z}\frac{\Deltac}{18\pi^2}\right)^{-1/2} \left(\frac{1+z}{10}\right)^{-3/2} \left(\frac{\Tvir}{10^4 \, \text{K}}\right)^{3/2} \, \Msun
\end{multline}
where $\mu$ is the mean molecular weight, $\Om^z=\Om(1+z)^3/( \Om(1+z)^3 + \OL + \Ok(1+z)^2)$, and $\Deltac=18\pi^2+82d-39d^2$ is the halo overdensity relative to the critical density at collapse with $d=\Om^z-1$.

\begin{figure}
  \centering
    \includegraphics[width=\linewidth]{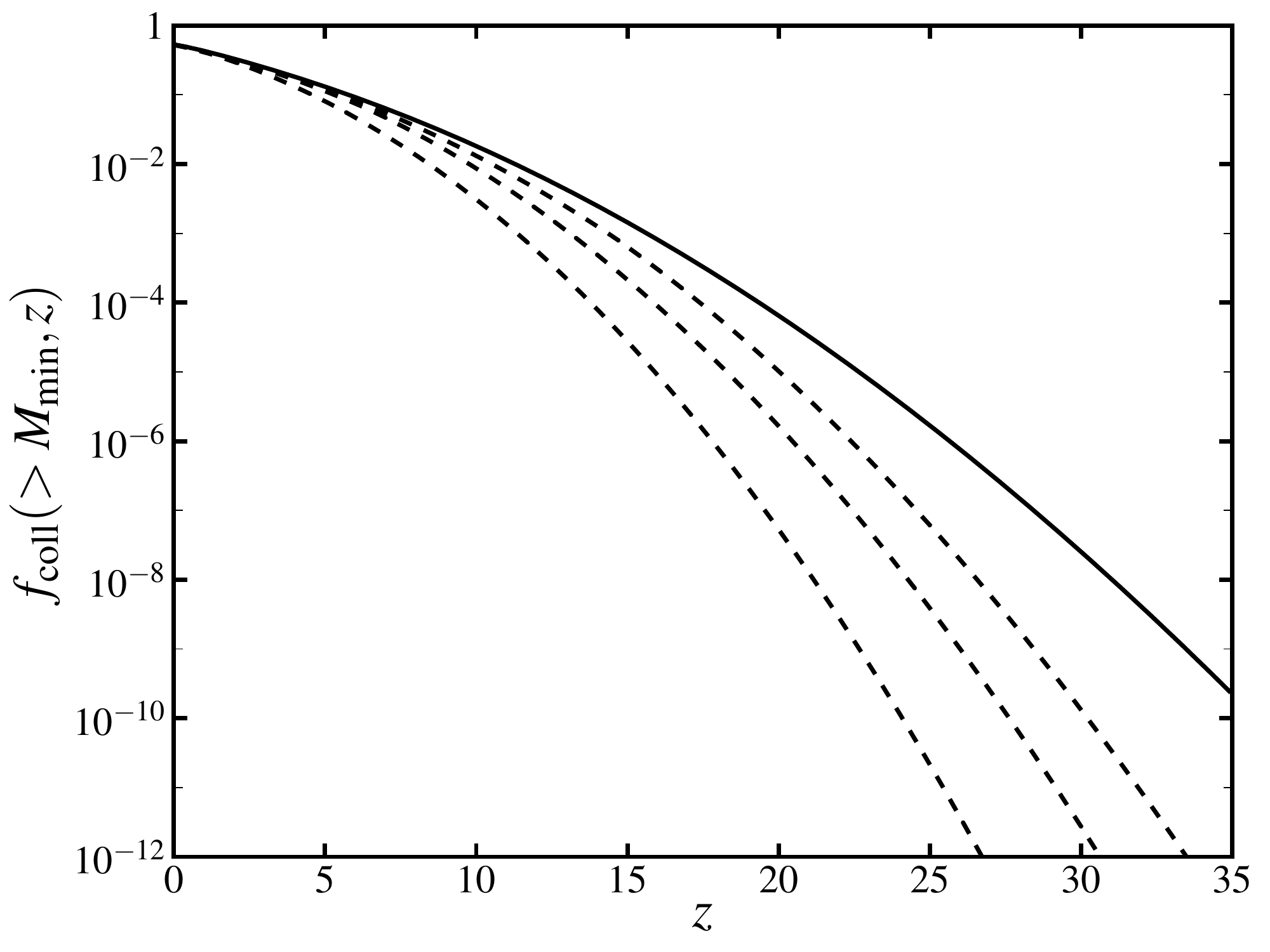}
	\caption{Mean collapse fraction for CDM (solid) and WDM (dashed) models. The WDM curves in ascending order are for $\mx=2,3,4 \, \text{keV}$. The collapse fraction is calculated using Eq.~(\ref{fcoll}) with $\Msf$ set by $T_{\text{vir}}=10^4 \,\text{K}$.}
\label{fcollmean}
\end{figure}

The mean collapse fraction in CDM and WDM models can be seen in Fig.~\ref{fcollmean}. At high redshifts, small halos begin to collapse in CDM, while no or few such halos collapse in WDM, resulting in a large relative difference between the collapse fractions in these models. However, this difference becomes smaller with lower redshifts as objects on scales larger than that inhibited by WDM start to collapse in both models. At late times, in the CDM scenario the mass within halos of sizes suppressed by WDM only represents a small fraction of the total mass within all collapsed structures, so the relative difference between the mean collapse fraction in CDM and WDM models is small at those times. Therefore, while structure formation is delayed in WDM models, the mean collapse fraction raises more rapidly as compared to CDM. 

\section{Cosmic 21-cm signal}
\label{21cmsignal}

The brightness temperature of the 21-cm signal measured against the CMB at redshift $z$ is given by
\begin{align}
\delta \Tb(z) &= \frac{\Ts-T_{\gamma}}{1+z} (1-e^{-\tau_{\nu_0}}) \nonumber \\
&\approx 27 x_{\text{HI}} (1+\delta)\left(1-\frac{T_{\gamma}}{\Ts}\right)  \left(\frac{1+z}{10}\frac{0.15}{\Om h^2}\right)^{1/2}  \nonumber \\ &\times\left(\frac{\Ob h^2}{0.023}\right) \left(\frac{H}{H+dv_{\parallel}/dr_{\parallel}}\right) \, \text{mK}
\end{align}
where $\tau_{\nu_0}$ is the optical depth at the 21-cm frequency $\nu_0$, $\Ts$ and $T_{\gamma}$ are the spin and CMB temperatures, respectively, $x_{\text{HI}}$ is the neutral fraction of hydrogen, $\delta$ is the overdensity, $H$ is the Hubble parameter and $dv_{\parallel}/dr_{\parallel}$ is the comoving velocity gradient along the line of sight. The spin temperature can be represented by
\begin{equation}
\Ts^{-1} = \frac{T_{\gamma}^{-1} + x_{\alpha}T_{\alpha}^{-1} + x_{c}T_{K}^{-1} }{1+x_{\alpha}+x_{c}} 
\end{equation}
where $\Tk$ and $T_{\alpha}$ are the kinetic and colour temperatures, respectively, and $x_c$ and $x_{\alpha}$ are the collisional and WF coupling coefficients, respectively. 

The earliest possible measurable cosmic 21-cm signal would be emitted during the `dark ages' before significant star formation occurs. At these early times, the gas is dense enough so that collisional coupling is strong and $\Ts\approx \Tk$. Before $z\sim150$, residual free electrons strongly couple the gas kinetic temperature to the CMB through Compton scattering, so $\Ts\approx \Tk \approx T_{\gamma}$ and no 21-cm signal can be observed at this time. After this point, any remaining free electrons are so defuse that the gas is decoupled from the CMB and cools adiabatically as $\Tk\propto (1+z)^2$. Since the CMB temperature decreases at the slower pace of $T_{\gamma}\propto (1+z)$, a 21-cm signal in absorption may be observed (at least in principle) at this time (\citealt{loeb2004,bharadwaj2004,naoz2005,lewis2007}). As the gas continues to cool, the collisional coupling becomes less efficient, driving $\Ts$ back up to the CMB temperature. As this scenario is relatively unaffected by structure formation, we do not expect the presence of WDM to significantly affect this era of the 21-cm signal and will restrict our attention to later times with redshifts below $z\sim 35$.\footnote{On the other hand, these early epochs may be affected by dark matter decay or annihilation (\citealt{mapelli2006, valdes2013}).}

It will be important to keep in mind that the kinetic temperature of the gas will be lower than the CMB temperature when WF coupling first becomes effective. As the Lyman-$\alpha$ background grows, the increasing strength of the WF coupling will drive $\Ts$ from a value near the CMB temperature to the lower kinetic temperature of the gas, thus producing another absorption signal. As WDM delays structure formation, the production of significant UV and X-ray backgrounds will be delayed, which in turn modifies the WF coupling, X-ray heating, and reionization. We therefore focus our attention to the astrophysical epochs in the 21-cm signal.

\section{Simulation of 21-cm signal}
\label{21cmmodel}

The 21-cm signal is simulated using the publicly available \cmfast \, code.\footnote{http://homepage.sns.it/mesinger/Download.html} This is a seminumerical simulation that generates density, velocity, ionization and spin temperature fields in a 3D box with length size $\sim$Gpc. In this section we briefly summarize the code. See \citealt{mesinger2011, mesinger2013b} and references within for further details.

An initial linear density field is generated as a Gaussian random field described by a power spectrum. The initial linear density field is then evolved using the Zel'Dovich approximation. 

Since we will be examining high-redshift eras, it will be necessary to compute the spin temperature and consequently the colour and kinetic temperatures and their associated coupling coefficients. The WF coupling coefficient $x_{\alpha}$ is given by
\begin{equation}
x_{\alpha} = S_{\alpha} \frac{J_{\alpha}}{J_{\nu}^c}
\end{equation}
where $J_{\alpha}$ is the angle-averaged Lyman-$\alpha$ background flux, $S_{\alpha}$ is a quantum correction term and $J_{\nu}^c=5.825\times10^{-12}(1+z) \,\text{cm}^{-2} \text{s}^{-1}\text{Hz}^{-1}\text{sr}^{-1}$. $S_{\alpha}$ and the colour temperature $T_{\alpha}$ are computed according to \citealt{hirata2006}. The kinetic temperature $\Tk$ is calculated by solving the set of (local) coupled differential equations for $\Tk$ and the ionized fraction $x_{\text{e}}$ in the neutral IGM, given by
\begin{subequations}\label{grp}
	\begin{equation}
		\frac{d x_{\text{e}}(\textbf{x},z)}{dz} = \frac{dt}{dz} \left( \Lambda_{\text{ion}} - \alpha_{\text{A}} C x_{\text{e}}^2 n_{\text{b}} f_{\text{H}} \right)
	\end{equation}
	\begin{multline}
		\frac{d\Tk(\textbf{x},z)}{dz} = \frac{2}{3k_{\text{b}}(1+x_{\text{e}})} \frac{dt}{dz} \sum_p \epsilon_p \\ + \frac{2\Tk}{3n_{\text{b}}}\frac{dn_{\text{b}}}{dz} - \frac{\Tk}{1+x_{\text{e}}}\frac{dx_{\text{e}}}{dz} 
	\end{multline}
\label{Tk}
\end{subequations}
\!\!where $\Lambda_{\text{ion}}$ is the ionization rate per baryon, $\alpha_{\text{A}}$ is the case-A recombination coefficient, $C$ is the clumping factor, $n_\text{b}$ is the total baryon number density, $f_{\text{H}}$ is the hydrogen number fraction, and $\epsilon_p$ is the heating rate for process $p$. The heating processes considered are X-ray heating $\epsilon_{\text{X}}$ and Compton heating $\epsilon_{\text{comp}}$.

The emission rate of photons at a particular frequency, which is needed to compute $\epsilon_{\text{X}}$, $\Lambda_{\text{ion}}$, and $J_{\alpha}$, is estimated by assuming that it is proportional to the star-formation rate, which is approximated using the growth of the collapse fraction. The comoving emissivity $e$ at frequency $\nu$ is then
\begin{equation}
e(\nu) = f_* \rho_{\text{b}} N(\nu) \frac{df_{\text{coll}}}{dt}
\end{equation}
where $f_*$ is the fraction of baryons that are incorporated into stars, $\rho_{\text{b}}=\bar{\rho}_{\text{b}}(1+\delta_{\text{nl}})$ is the total baryon density including the non-linear overdensity $\delta_{\text{nl}}$, and $N(\nu)$ is the number of photons with frequency $\nu$ per solar mass in stars. The local collapse fraction is computed using the hybrid prescription of \citealt{barkana2004}, where the biased EPS method is used to compute relative local halo abundances whose mean is then normalized to fit the mean collapse fraction given by the Sheth-Tormen mass function in Eq.~(\ref{fcoll}).

Ionization fields are generated by assuming that a region is ionized if it contains more ionizing photons than neutral hydrogen atoms (multiplied by $1+\bar{n}_{\text{rec}}$, where $\bar{n}_{\text{rec}}$ is the mean number of recombinations per baryon). The excursion-set formalism is used with the condition that $\zeta f_{\text{coll}}(\textbf{x},z,R) \geq 1 - x_{\text{e}}(\textbf{x},z,R) $ for a cell centred at location $\textbf{x}$ to be fully ionized, where $f_{\text{coll}}(\textbf{x},z,R)$ is the collapse fraction smoothed on scale $R$, $\zeta$ is the ionization efficiency, and $1-x_{\text{e}}(\textbf{x},z,R) $ is the remaining fraction of neutral hydrogen within $R$. This criteria is evaluated at deceasing scales $R$ and if the cell is not marked as fully ionized as the scale of the pixel length is reached, the cell's ionization fraction is marked as $\zeta f_{\text{coll}}(\textbf{x},z,R_{\text{cell}}) + x_{\text{e}}(\textbf{x},z) $. Lastly, we note that the ionization efficiency can be decomposed as $\zeta=A_{{\text{He}}}f_*f_{\text{esc}}N_{\text{ion}}/(1+\bar{n}_{\text{rec}})$, where $f_{\text{esc}}$ is the fraction of ionizing photons that escape their host galaxy, $N_{\text{ion}}$ is the number of ionizing photons per baryon inside stars and $A_{\text{He}}$ is a correction factor due to the presence of Helium.

\section{Simulation Results}
\label{results}

As much is unknown about astrophysical properties during high-redshift eras, we will examine possible degeneracies in the 21-cm signal between WDM and astrophysical quantities. As a first step, we will compare the delayed WDM 21-cm signal with that in CDM with a reduced photon-production efficiency. Specifically, we decrease the efficiency uniformly over frequency by decreasing $f_*$, but note that $f_*$ is degenerate with other parameters used to calculate photon production efficiencies.

The box used in our simulation runs was 750 Mpc on a side and was comprised of $300^3$ cells. The 21-cm signal was simulated in the redshift range $z = 5.6$ to $35$. We set the minimum halo virial temperature that supports star formation to be $T_{\text{vir}}=10^4 \, \text{K}$ as to approximate the minimum temperature need to efficiently cool the halo gas through atomic cooling, neglecting possible feedback processes.\footnote{Although the very first stars were likely formed within smaller halos with $T_{\text{vir}}$  on the order of $10^3 \, \text{K}$ that were molecularly cooled, star formation in such halos can easily be disrupted by feedback processes (\citealt{haiman2000, mesinger2009}) and we therefore neglect radiation from sources located in such halos.} Our fiducial model uses a $f_*$ value of $f_{*\text{fid}}=10\%$ and an ionization efficiency $\zeta=31.5$.

\begin{figure}
  \centering
    \includegraphics[width=\linewidth]{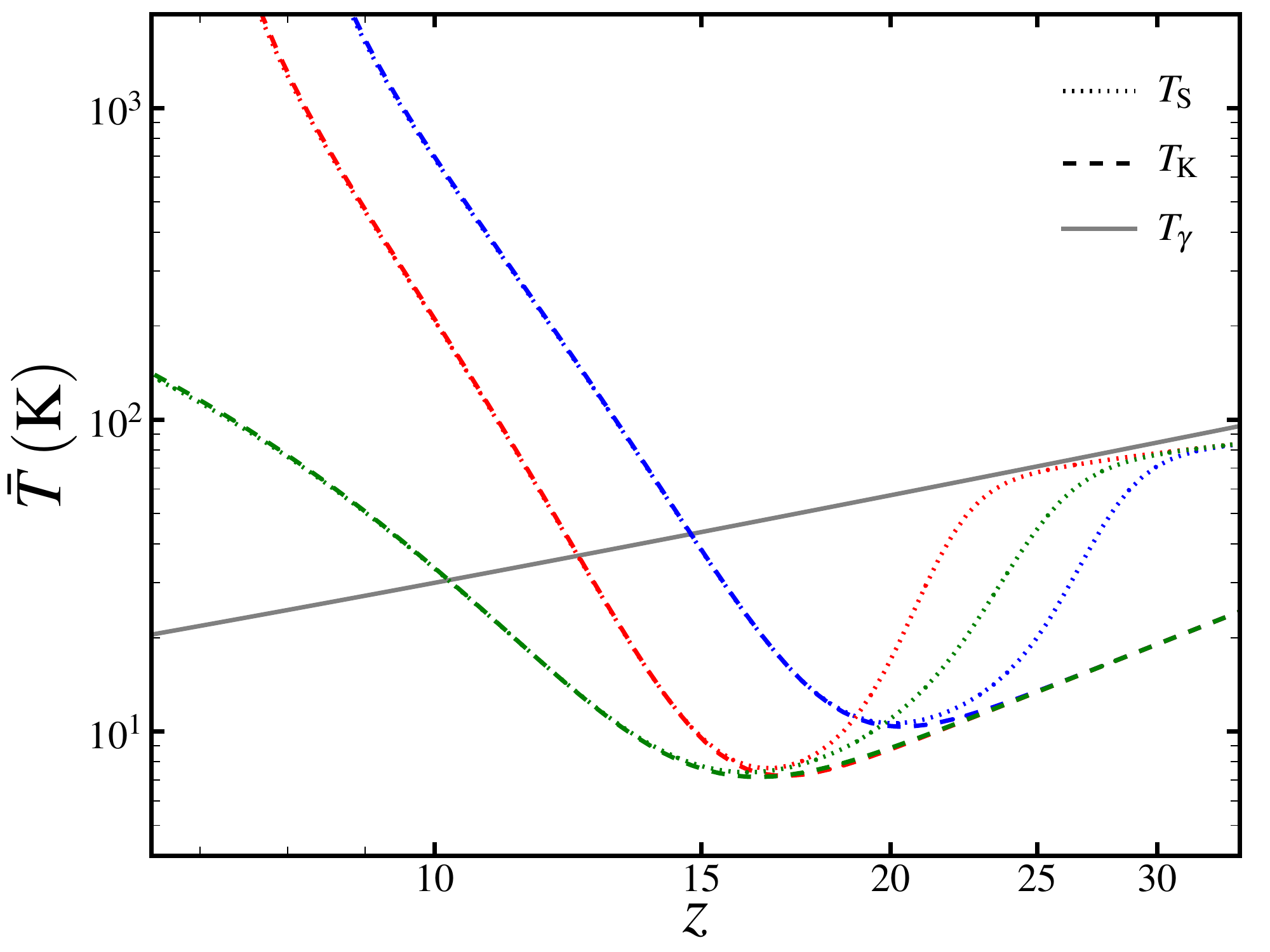}
	\caption{Mean spin temperatures $\bar{T}_{\text{S}}$ for CDM and WDM models. The dotted curves show $\bar{T}_{\text{S}}$ for our fiducial CDM model (blue), WDM with $\mx=3\,\text{keV}$ (red), and CDM with $f_*/f_{*\text{fid}}=0.1$ (green). In addition, the mean kinetic temperature $\bar{T}_{\text{K}}$ of each model is plotted with a dashed curve in the same colour used for $\bar{T}_{\text{S}}$. The grey solid line is the CMB temperature.}
\label{Ts}
\end{figure}

Examples of the mean spin and kinetic temperatures for CDM and WDM models are plotted in Fig.~\ref{Ts}. As expected, for WDM $\Ts$ stays near $T_{\gamma}$ for a longer time and the lowest point in the absorption trough, where the X-ray heating rate first surpasses the adiabatic cooling rate, occurs later. As mentioned in Section~\ref{halos}, although the mean collapse fraction is lower in WDM models, it grows more rapidly, which is reflected in the heating of the gas. In addition, Fig.~\ref{Ts} shows curves for CDM with the lower $f_*$ value of $f_*/f_{*\text{fid}}=0.1$, which in our model happens to delay star formation such that the minimum value of $\bar{T}_{\text{S}}$ occurs roughly at the same time as in the WDM example used. In this case, the X-ray heating rate increases at a much slower rate after the minimum in $\bar{T}_{\text{S}}$ as compared to the two other cases shown, since lowering $f_*$ reduces the photon production efficiency in stars of all masses. In both non-fiducial cases shown, $\bar{T}_{\text{S}}$ and thus $\delta \bar{ T}_{\text{b}}$ reach a lower value in their absorption troughs since the gas undergoes further cooling in the extra time needed for the X-ray heating to become efficient.

\begin{figure*}
\centering
\subfloat[]{\label{Tbave}\includegraphics[width=0.5\linewidth]{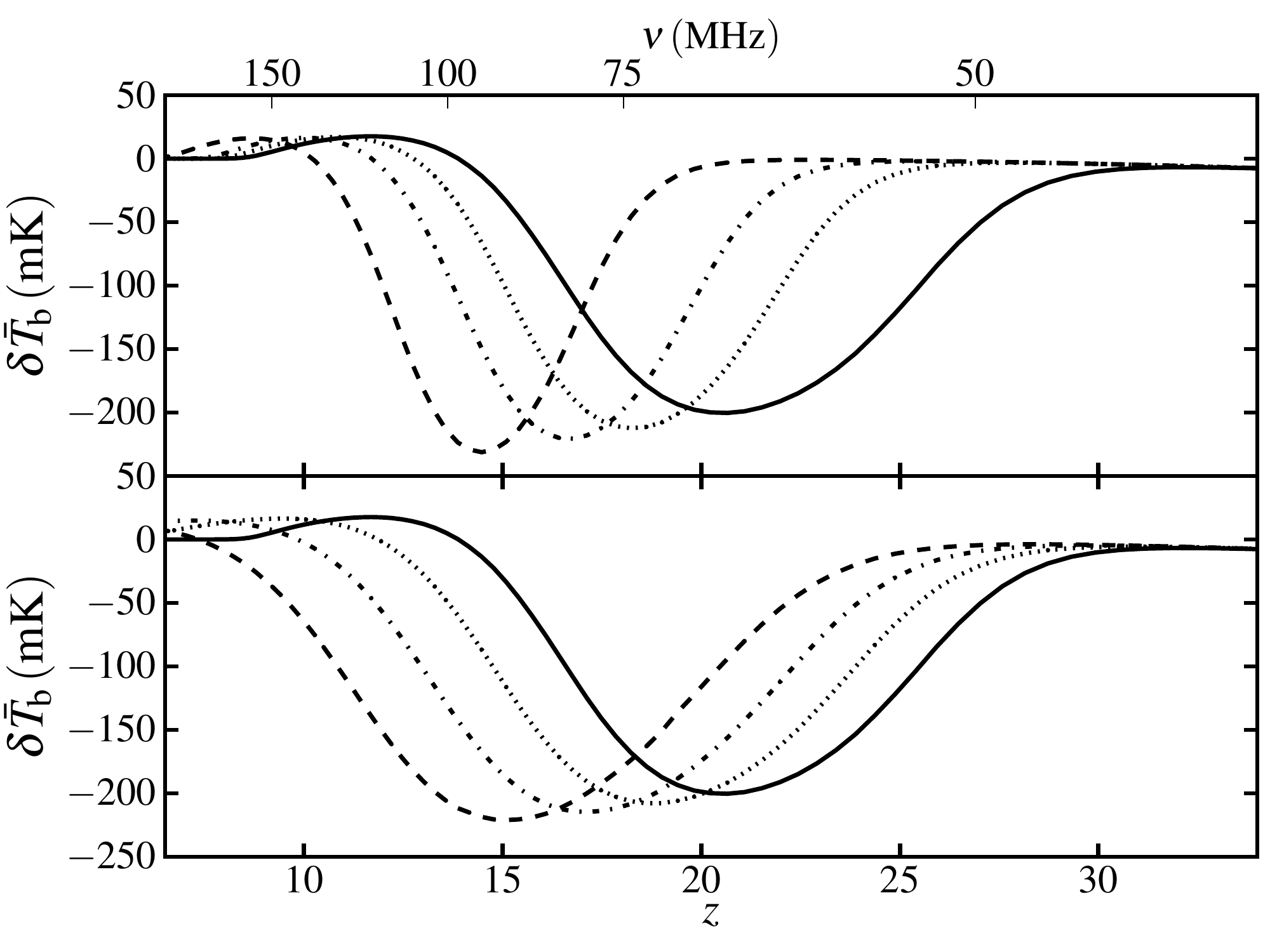}}
\subfloat[]{\label{dTb}\includegraphics[width=0.5\linewidth]{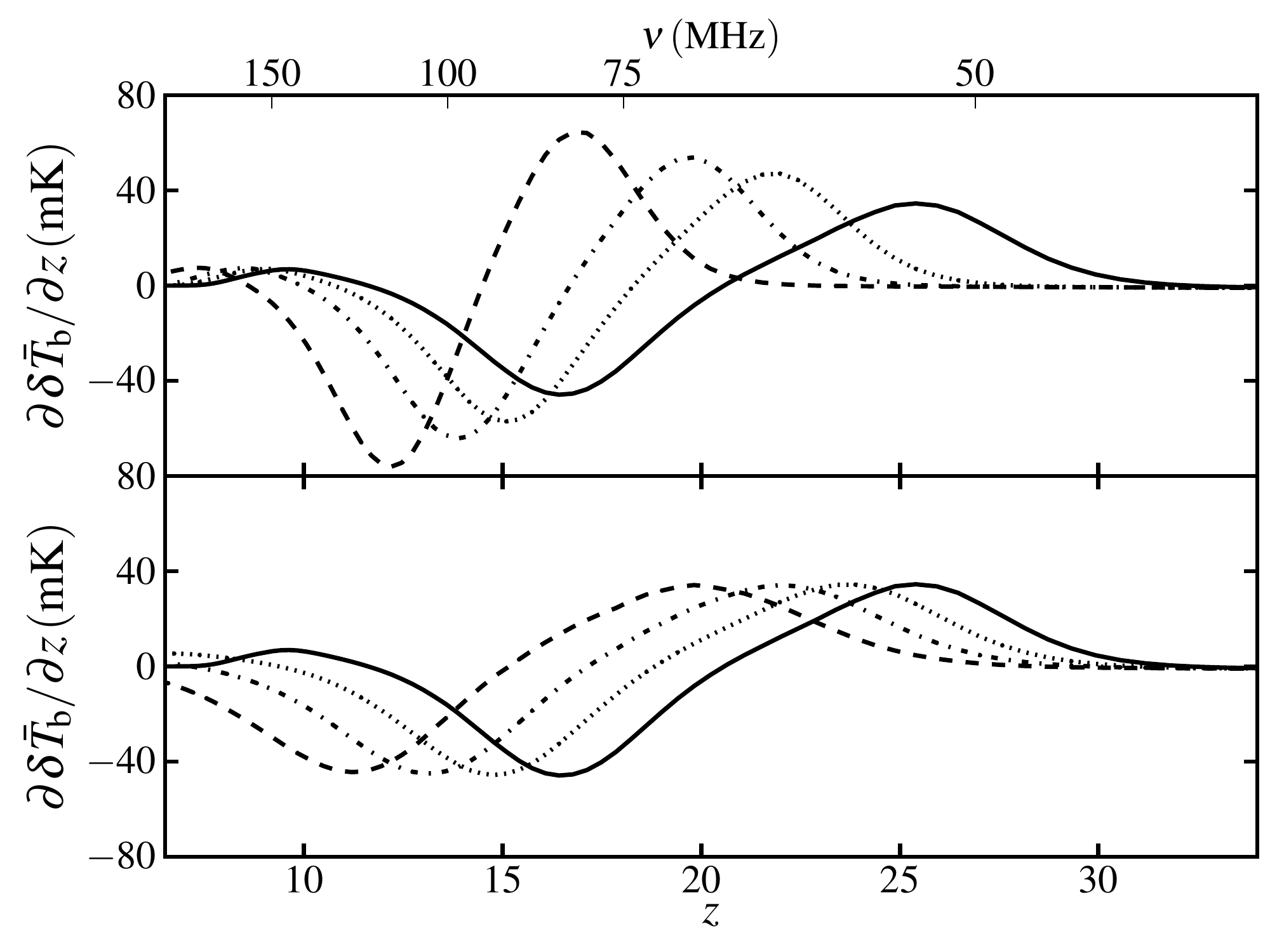}}
\caption{Mean 21-cm brightness temperature $\delta \bar{ T}_{\text{b}}$ (a) and its derivative with respect to redshift (b). In all plots, the solid curve is the fiducial CDM model. The upper plots show the results of WDM runs where the dashed, dotted-dashed, and dotted curves are for $\mx=2,3,4 \, \text{keV}$, respectively. The lower plots show CDM runs where the dashed, dotted-dashed, and dotted curves are for CDM models with $f_*/f_{*\text{fid}}=0.03,0.1,0.5$, respectively.}
\label{Tb}
\end{figure*}

The evolution of the mean brightness temperatures for WDM models with $\mx=2,3,4\,\text{keV}$ are shown in Fig.~\ref{Tb}. \footnote{We caution the reader that WDM models with $\mx=2,3\,\text{keV}$ are disfavoured by recent Lyman-$\alpha$ observations (\citealt{viel2013}). However, Lyman-$\alpha$ forest constraints are still susceptible to astrophysical (thermal and ionization history) and observational (sky and continuum subtraction) degeneracies. Therefore, it is still useful to confirm these constraints using the redshifted 21-cm signal.} It is readily seen that having WDM with a particle mass of a few keV can substantially change the mean 21-cm brightness temperature evolution. While lowering $f_*$ within CDM models can delay the strong absorption signal, the resulting absorption trough is much wider than in WDM. For the same delay in the minimum of $\delta \bar{ T}_{\text{b}}$, the delay in reionization is greater for CDM than for WDM. Although reionization may be greatly delayed, well past $z=6$, in models with low values of $f_*$, our primary focus is on the pre-reionization 21-cm signal. We caution against automatically discarding these models, as the star-formation efficiency may diverge from earlier values by reionization.

Examining the gradient of the global signal in Fig.~\ref{dTb}, we see the suppressing $f_*$ in CDM models only shifts the mean signal to lower redshifts. On the other hand, decreasing $\mx$ in WDM models increases the gradients of the mean signal. In CDM models, $\partial \delta \bar{ T}_{\text{b}} / \partial z$ attains values near $33 \, \text{mK}$ ($-45 \, \text{mK}$) near its maximum (minimum) regardless of its $f_*$ value. This can increase significantly in WDM models, for example to $\sim 64 \, \text{mK}$ ($\sim -77 \, \text{mK}$) at its maximum (minimum) for WDM with $\mx=2 \, \text{keV}$. 

\begin{figure*}
\centering
\subfloat[]{\label{zpts}\includegraphics[width=0.5\linewidth]{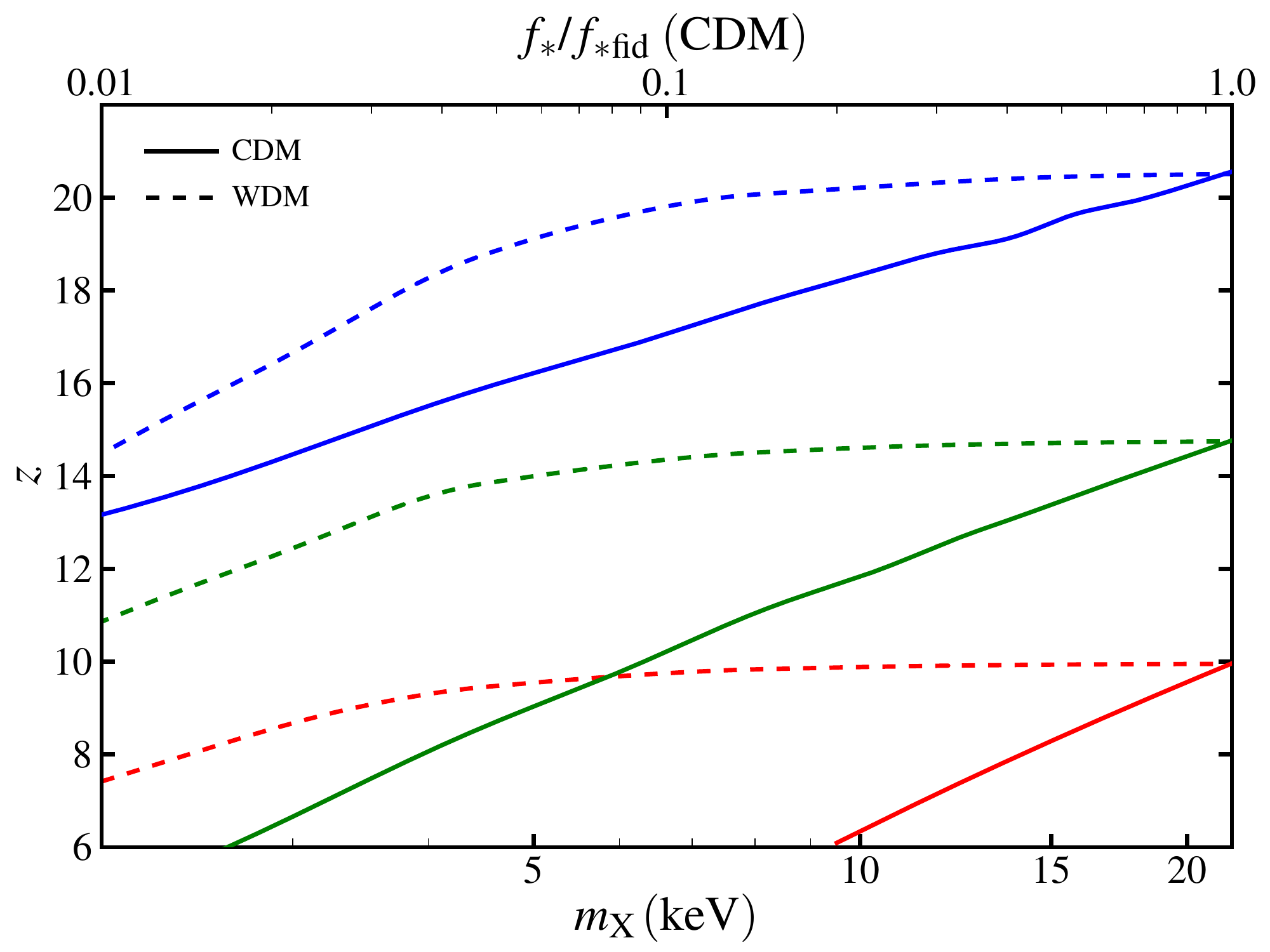}}
\subfloat[]{\label{degen}\includegraphics[width=0.5\linewidth]{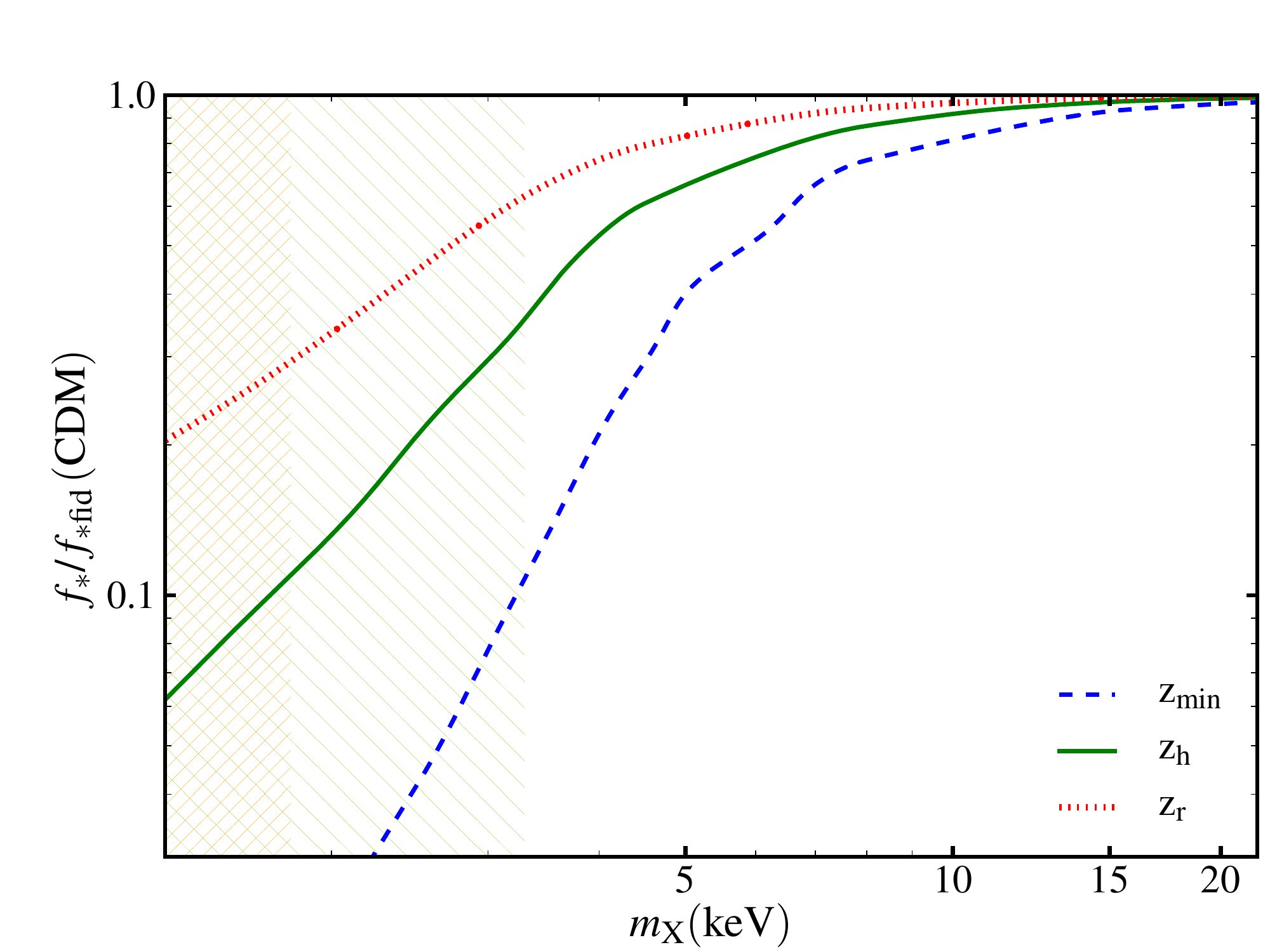}}
\caption{`Critical points' in the mean 21-cm signal. (a) Redshifts of critical points for CDM (solid curves) and WDM (dashed curves) models. For CDM curves, the redshifts of the critical points are plotted as a function of $f_*$, which can be read from the top horizontal axis. For WDM curves, the critical point redshifts are plotted as a function of $\mx$, the values of which can be read from the lower horizontal axis. In descending order from the right, the curves are the redshifts $\zmin$ (blue), $\zh$ (green), and $\zr$ (red) for each model. (b) Parameter space curves $z_e(f_*|\text{CDM})=z_e(\mx|\text{WDM})$ for various critical points $z_e \in \{ \zmin, \zh, \zr \}$. The orange (green) hatched region shows models disfavoured by observations of GRBs (the Lyman-$\alpha$ forest) from \citealt{desouza2013} (\citealt{viel2013}).  }
\end{figure*}

The effect of WDM on the global 21-cm signal can be tracked through different `critical points' in the signal's evolution. We choose these points to be the redshift $\zmin$ at which $\delta \bar{ T}_{\text{b}}$ reaches its minimum value, the redshift $\zh$ when the kinetic temperature of the gas is heated above the CMB temperature, and the redshift of reionization $\zr$ taken to be the redshift where the mean ionized fraction is $\bar{x}_i(\zr)=0.5$. These points are plotted for both CDM and WDM in Fig.~\ref{zpts}. The solid curves track the effect of lowering $f_*$ on the redshifts of the critical points in CDM models (the values of $f_*$ can be read from the upper horizontal axis). The dashed curves show the effect of WDM on these redshifts, where the value of $\mx$ for each model can be read from the lower horizontal axis. 

We begin to explore possible degeneracies between CDM and WDM cosmologies by finding the value of $f_*$ required in CDM that would have a particular critical point occur at the same redshift as it would in WDM with a particular value of $\mx$. In other words, for a particular event that occurs at redshift $z_e$, we would like to find the curve that satisfies $z_e(f_*|\text{CDM})=z_e(\mx|\text{WDM})$. These curves for $\zmin$, $\zh$, and $\zr$ can be seen in Fig.~\ref{degen}. We can see that if one uses the milestone $\zr$ to distinguish between CDM and WDM with $\mx=2,3,4 \, \text{keV}$ then $f_*$ has to be known within a factor of 3.0, 1.8, and 1.4, respectively. Using $\zmin$ instead, $f_*$ only has to be known within a factor of 50, 13, and 4.8 for $\mx=2,3,4 \, \text{keV}$, respectively, since the impact of WDM is larger at higher redshifts. Near $\mx=15 \, \text{keV}$, using $\zmin$ to distinguish WDM from CDM requires $f_*$ to be known within a factor of 1.1 and drops to 1.01 by $\mx\sim 20 \, \text{keV}$ (although the astrophysical motivations for WDM as mentioned in the introduction loses much of its appeal past a few keV).

As the value of $\mx$ is lowered, the curves in Fig.~\ref{degen} diverge from one another, as the more rapid growth of structure in WDM changes the relative timing of the milestones. Therefore, if $f_*$ is approximately constant throughout the epochs under consideration, adjusting the value of $f_*$ in CDM so that a particular critical point occurs at the same redshift as it does in WDM will misalign other critical points and thus cannot reproduce the whole history of $\delta \bar{ T}_{\text{b}}$ in WDM models. 

\begin{figure}
  \centering
    \includegraphics[width=\linewidth]{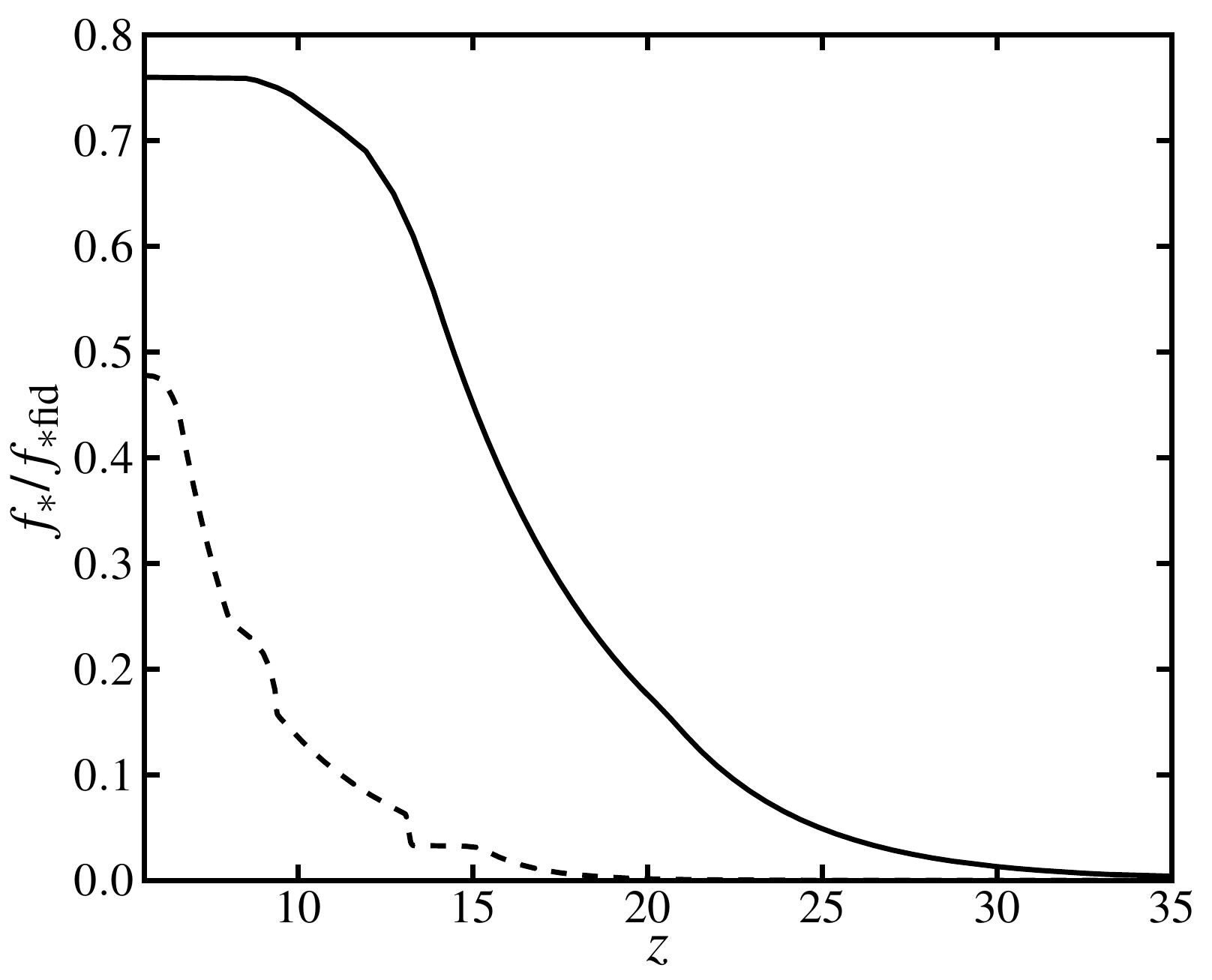}
	\caption{Evolution of $f_*(z)$ in CDM required to match the mean brightness temperature $\delta \bar{ T}_{\text{b}}$ in WDM with $\mx=2  \, \text{keV}$ (dashed) and $\mx=4  \, \text{keV}$ (solid). All other parameters are set to their values in the fiducial CDM model.}
\label{fstarz}
\end{figure}

However, we can mimic the WDM mean brightness temperature evolution with CDM if we allow $f_*$ to vary in time. To illustrate this, Fig.~\ref{fstarz} shows the form of $f_*(z)$ needed to reproduce the mean 21-cm signal for WDM with $\mx=2,4  \, \text{keV}$. At high redshifts ($z\gtrsim15,25$ for $\mx=2,4  \, \text{keV}$), $f_*$ is more than an order of magnitude smaller than its value at the end of reionization to compensate for the delay of structure formation in WDM. When more massive halos start to collapse (near $z=10,20$ for $\mx=2,4  \, \text{keV}$), $f_*$ rises quickly by roughly an order of magnitude to mimic the more rapid change of the collapse fraction in WDM and finally levels off during reionization. While this evolution of $f_*$ may be possible, it seems contrived without an underlying model of such evolution.

\begin{figure*}
\centering
\subfloat[]{\includegraphics[width=0.5\linewidth]{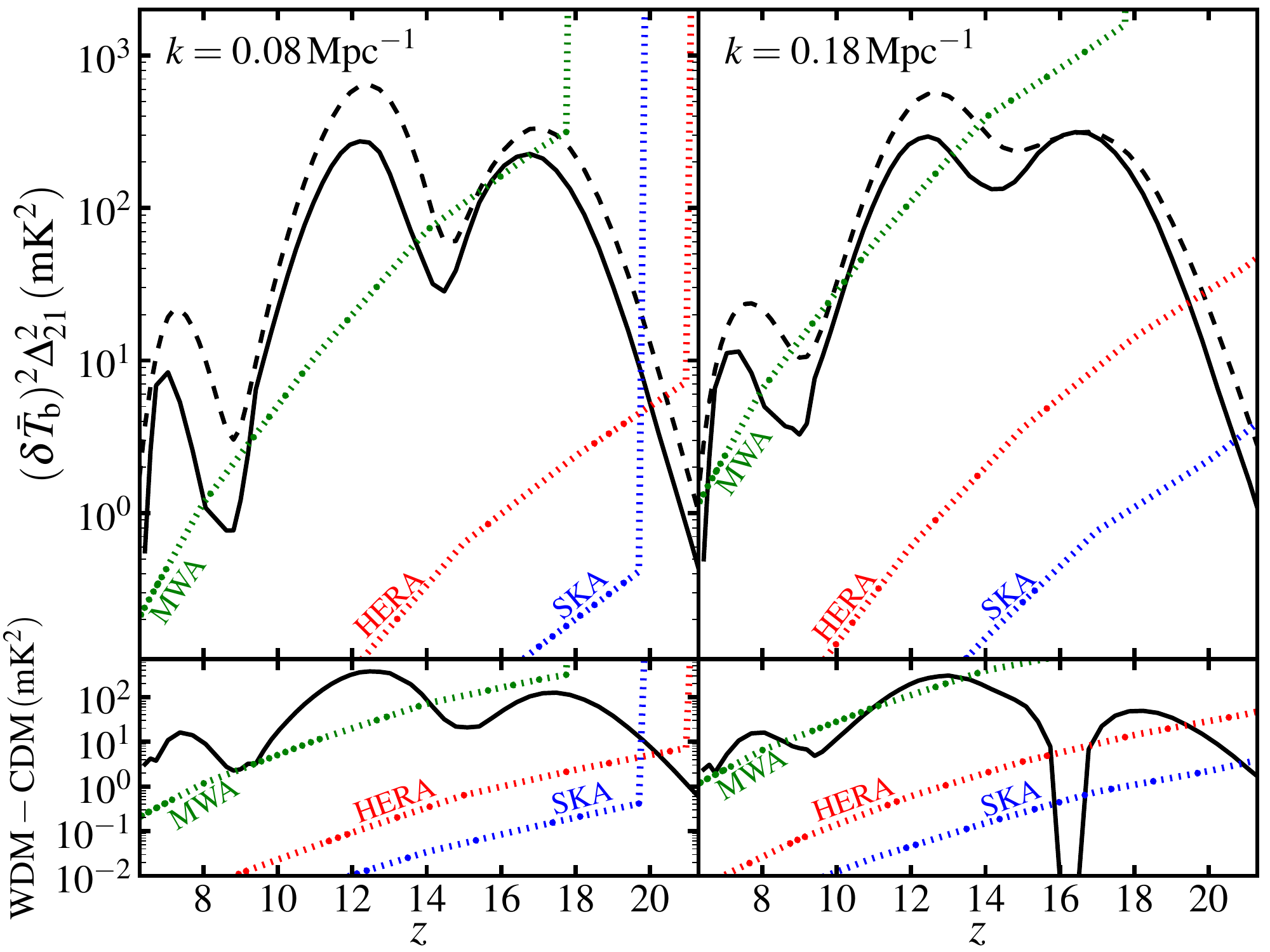}}
\subfloat[]{\includegraphics[width=0.5\linewidth]{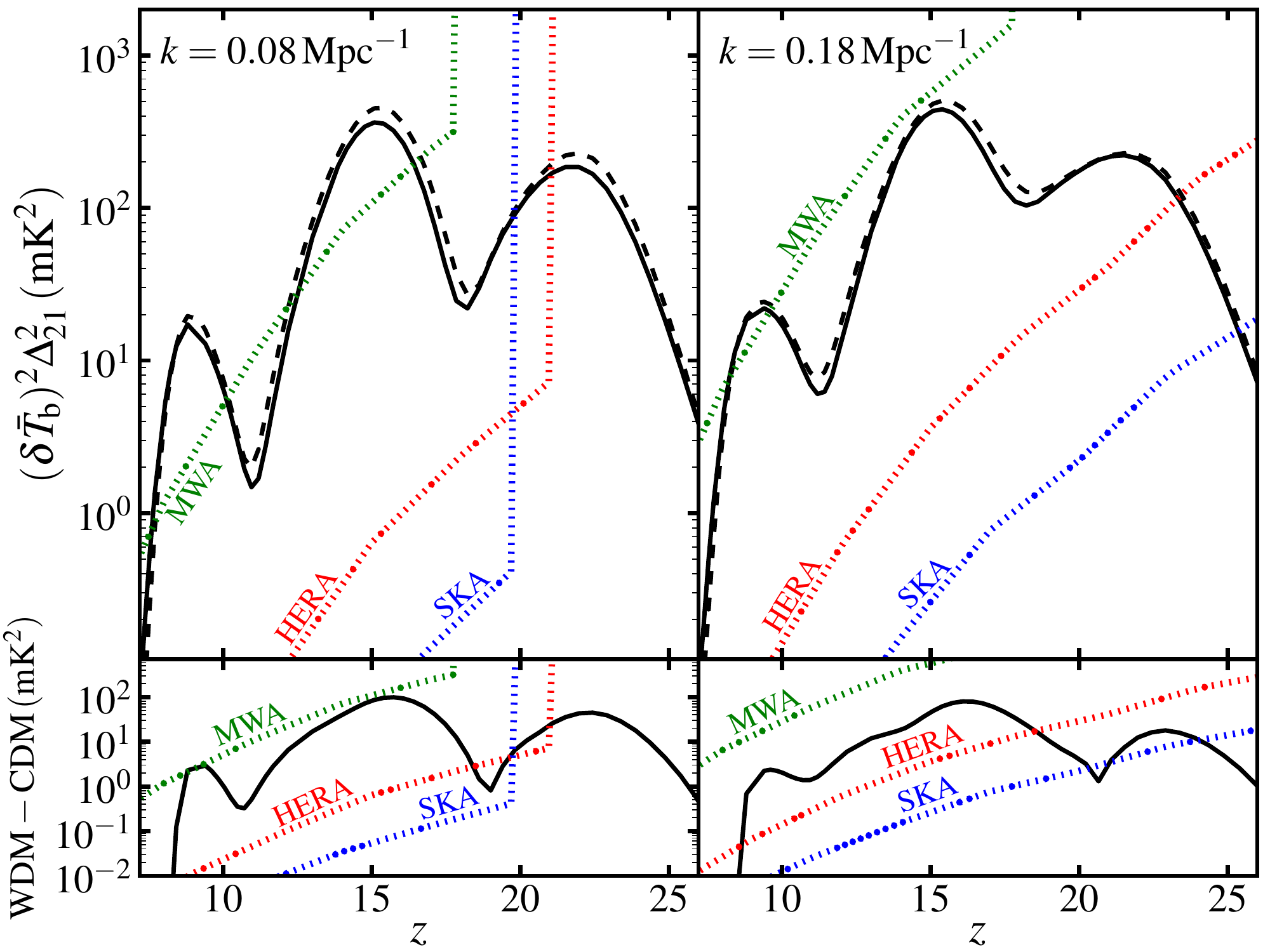}}
\caption{Evolution of the power spectrum of $\delta \Tb$ for WDM with (a) $\mx=2 \, \text{keV}$ and (b) $\mx=4 \, \text{keV}$. The top panels show power spectra at $k=0.08, 0.18 \, \text{Mpc}^{-1}$ for WDM (dashed) and the CDM model (solid). CDM models have $f_*(z)$ chosen to reproduce the global 21-cm signal found for the respective WDM model. The bottom panels show the difference in the power spectrum between WDM and CDM models. Dotted curves show forecasts for the $1-\sigma$ power spectrum thermal noise as computed in \citealt{mesinger2013a} with 2000h of observation time. The dotted green, blue, and red curves are the forecasts for the MWA, SKA, and HERA, respectively.} 
\label{ps_k}
\end{figure*}

Even in cases where $f_*$ evolves in such a way as to mimic the mean brightness temperature in WDM, one can differentiate between WDM and CDM by examining the spectrum of perturbations in the 21-cm signal at certain points in its evolution. Perturbations in the UV and X-ray fields add power to the 21-cm power spectrum $\Delta_{21}^2$ on large scales. Since the bias of sources in WDM can be greater than that in CDM (\citealt{smith2011}), more power is added on large scales in WDM than in CDM. This effect is most easily seen at times when inhomogeneities in $x_{\alpha}$ or $\Tk$ are at their maximum. Fig.~\ref{ps_k} shows the evolution of the power spectrum for the modes $k=0.08 \, \text{Mpc}^{-1}$ and $k=0.18 \, \text{Mpc}^{-1}$, showing a three peak structure, where the peaks from high to low redshift are associated with inhomogeneities in $x_{\alpha}$, $\Tk$, and $x_{\text{HI}}$, respectively. When inhomogeneities in $\Tk$ are at their maximum, the power at  $k=0.08,0.18 \, \text{Mpc}^{-1}$ can be boosted in WDM by as much as a factor of $2.4,2.0 \, (1.3,1.1)$ for $\mx=2 \, \text{keV}$ ($\mx=4 \, \text{keV}$). When inhomogeneous in $x_{\alpha}$ are near their height, the power at $k=0.08 \, \text{Mpc}^{-1}$ can be increased by a factor of 1.5 (1.2) for WDM with $\mx=2 \, \text{keV}$ ($\mx=4 \, \text{keV}$). 

Current and next generation interferometric radio telescopes may be used to detect the boost in power associated with WDM models. The dotted curves in Fig.~\ref{ps_k} show forecasts for the $1-\sigma$ power spectrum thermal noise levels for 2000 hours of observation time, computed by \citealt{mesinger2013a}, for the Murchison Widefield Array (MWA)\footnote{http://www.mwatelescope.org/}, the Square Kilometre Array (SKA)\footnote{http://www.skatelescope.org/}, and for the proposed Hydrogen Epoch of Reionization Array (HERA)\footnote{http://reionization.org}. This estimate is quite conservative in that it ignores the contribution of foreground-contaminated modes (\citealt{pober2013}). From these forecasts, we can see that the MWA may be able to at least marginally detect the boost in power for the $\mx=2\,\text{keV}$ model at the reionization and X-ray heating peaks. In addition, these estimates indicate that next generation instruments will be able to easily measure the excess of power at these scales for $\mx=2,4 \, \text{keV}$ models over a wide range of redshifts.

\begin{figure}
  \centering
    \includegraphics[width=\linewidth]{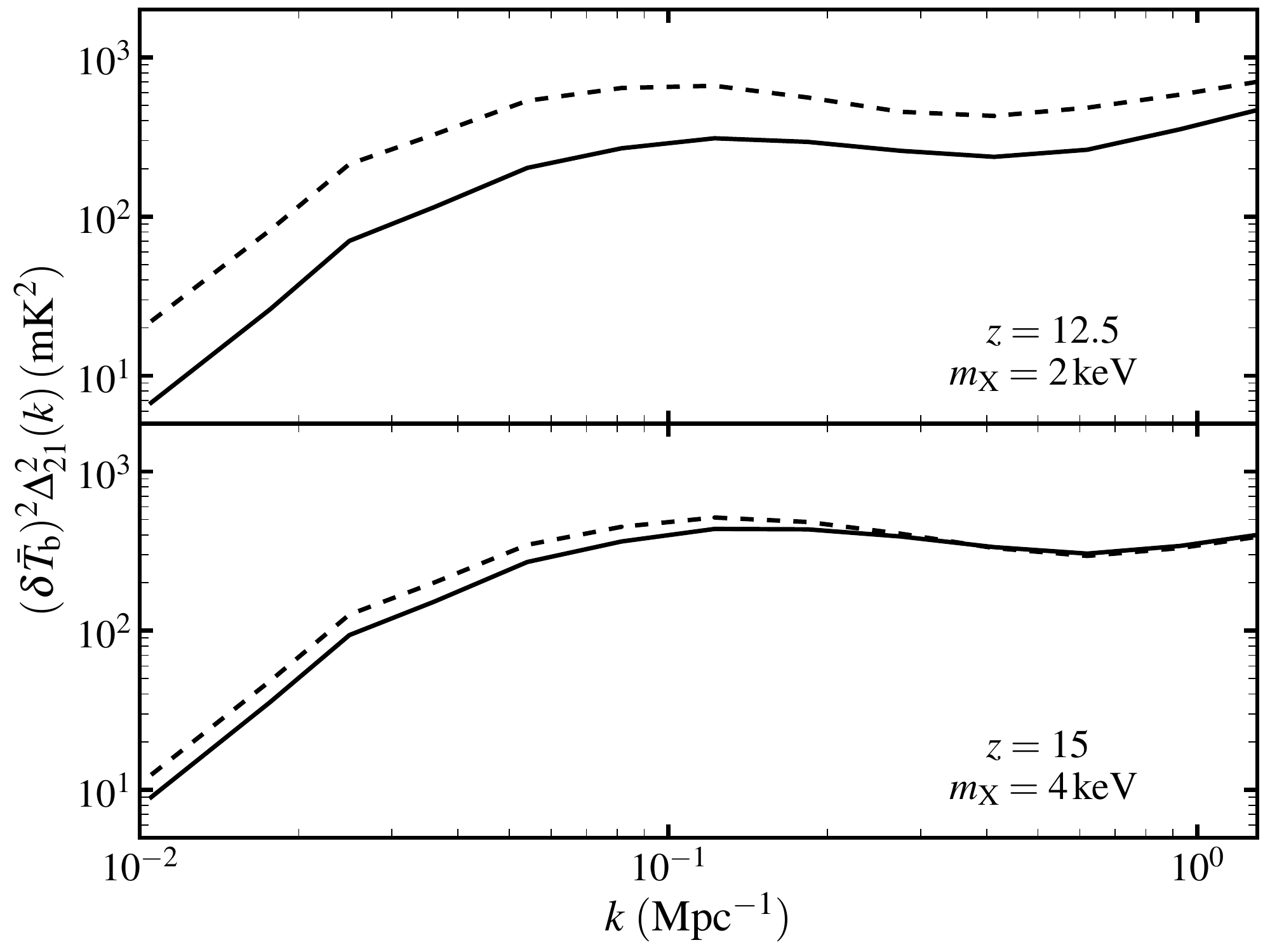}
	\caption{Power spectrum of the brightness temperature $\delta \Tb$. The top panel shows the power spectrum at $z=12.5$ for WDM with $\mx=2  \, \text{keV}$ (dashed) and CDM (solid). In the CDM model, $f_*(z)$ evolves as shown in Fig.~\ref{fstarz} such that it reproduces the global signal in the WDM model. Similarly, the bottom panel shows the power spectrum at $z=15$ for WDM with $\mx=4  \, \text{keV}$ (dashed) and CDM (solid) with $f_*(z)$ chosen to match the global signal in this WDM model. The power spectrum of each model is plotted at a redshift near where the X-ray background is at its most inhomogeneous state in its respective model.}
\label{ps_z}
\end{figure}

The 21-cm power spectrum during a redshift near the time when $\Tk$ is at its most inhomogeneous state is plotted in Fig.~\ref{ps_z} for WDM with $\mx=2,4  \, \text{keV}$ and their CDM counterparts. One can see that the boost in power in WDM may continue to $k$ values lower than those used in Fig.~\ref{ps_k}. In particular, the power near $k=0.01 \, \text{Mpc}^{-1}$ in WDM models with $\mx=2 \, \text{keV}$ ($\mx=4 \, \text{keV}$) may be larger by a factor of 3 (1.3) as compared to in CDM models at these times.

Finally, we mention that for simplicity we have chosen to vary only one astrophysical property. By allowing other astrophysical parameters to vary as a function of redshift, most notably $\Mmin$, it might be possible to produce a 21-cm power spectrum degenerate with WDM throughout the redshifts under investigation and we leave this question for future work.

\section{Conclusions}

In warm dark matter models, the abundance of small halos is suppressed, which can leave a strong imprint at high redshifts. Since structure formation is delayed but more rapid in WDM, the mean 21-cm signal will follow suit, resulting in a delayed, deeper and more narrow absorption trough. These effects can easily be seen in the global 21-cm signal for WDM with free-streaming lengths above current observational bounds for thermal relic masses as high as $\mx \sim 10-20 \, \text{keV}$ ($\Rc \sim 6-13 \, \text{kpc}$). 

Suppressing the photon-production efficiency of astrophysical sources can delay the 21-cm signal as well. As such, to discriminate between WDM and CDM models by measuring the redshift of reionization, the photon-production efficiency must be known to within a factor of 3.0, 1.8, and 1.4 for WDM with $\mx=2,3,4 \, \text{keV}$ ($\Rc\approx 86,54,39  \, \text{kpc}$), respectively. Since the impact of WDM is larger at higher redshifts, if milestones in the mean 21-cm signal that occur at higher redshift are used to differentiate WDM and CDM models, the precision to which this efficiency must be known decreases. For example, if measuring the redshift of the minimum of the mean 21-cm signal (during the astrophysical epoch of the signal) the efficiency must only be known within a factor of 50, 13, and 4.8 for $\mx=2,3,4 \, \text{keV}$, respectively. 

If the star-formation remains approximately constant over the range of redshifts under consideration, degeneracy between CDM and WDM models may be broken by examining the gradient of the mean 21-cm signal, which is larger in WDM due to its more rapid pace of structure formation. In addition, the spectrum of perturbations in the 21-cm signal may as well be used to break this degeneracy, as the 21-cm power spectrum in WDM has an excess of power on large scales owing to the stronger biasing of sources in WDM.  This is true even if the photon-production efficiency evolves with redshift in such a way as to reproduce with CDM the global 21-cm signal in WDM models. For WDM with $\mx=2\, \text{keV}$ ($\mx=4\, \text{keV}$), the power in the 21-cm signal at $k = 0.08, 0.18 \, \text{Mpc}^{-1}$ can be increased by a factor as high as $2.4,2.0 \, (1.3,1.1)$ as compared to that in CDM. Power spectrum measurements made by current interferometric telescopes, such as the MWA, should be able to discriminate between CDM and WDM models with $\mx\lesssim 3\,\text{keV}$, while next generation telescopes will easily be able differentiate between CDM and all relevant WDM models.

In this work, we assume that atomically-cooled halos drive the 21-cm signal. If instead smaller, molecularly-cooled halos, whose production is suppressed in WDM, play a significant role in producing the 21-cm signal in CDM, then the effects differentiating WDM from CDM described above would be even more pronounced. On the other hand, if star-formation was not efficient in halos with $T_{\text{vir}}=10^4 \,\text{K}$, the differences between CDM and WDM in the 21-cm signal would be diminished.

\vskip+0.5in

\section*{Acknowledgements}

This research was supported in part by the National Science and Engineering Research Council of Canada (NSERC). MS is supported in part by a NSERC Canada Graduate Scholarship. The research of KS is supported in part by a NSERC Discovery Grant. KS thanks the Aspen Center for Physics, where part of this work was completed, for their hospitality. YZM is supported by a CITA National Fellowship.


\begin{thebibliography}{9}

\bibitem[\protect\citeauthoryear{Abazajian et al.}{2001}]{abazajian2001}
Abazajian, K., Fuller, G. M., \& Patel, M., 2001, Phys. Rev. D, 64, 023501

\bibitem[\protect\citeauthoryear{Barkana et al.}{2001}]{barkana2001}
Barkana, R., Haiman, Z., \& Ostriker, J. P., 2001, ApJ, 558, 482

\bibitem[\protect\citeauthoryear{Barkana \& Loeb}{2001}]{barkanaloeb2001}
Barkana, R., \& Loeb, A., 2001, Phys. Rep., 349, 125

\bibitem[\protect\citeauthoryear{Barkana \& Loeb}{2004}]{barkana2004}
Barkana R., \& Loeb A., 2004, ApJ, 609, 474

\bibitem[\protect\citeauthoryear{Bharadwaj \& Ali}{2004}]{bharadwaj2004}
Bharadwaj, S., \& Ali, S. S., 2004, MNRAS, 352, 142

\bibitem[\protect\citeauthoryear{Bode et al.}{2001}]{bode2001}
Bode, P., Ostriker, J. P., \& Turok, N., 2001,  ApJ, 556, 93

\bibitem[\protect\citeauthoryear{Bond et al.}{1982}]{bond1982}
Bond, J. R., Szalay, A. S., \& Turner, M. S., 1982, Phys. Rev. Lett., 48, 1636

\bibitem[\protect\citeauthoryear{Boyarsky et al.}{2009}]{boyarsky2009}
Boyarsky, A., Lesgourgues, J., Ruchayskiy, O., \& Viel, M., 2009, Phys. Rev. Lett., 102, 201304

\bibitem[\protect\citeauthoryear{Boylan--Kolchin et al.}{2011}]{boylan2011}
Boylan--Kolchin, M., Bullock, J. S., \& Kaplinghat, M., 2011, MNRAS, 415, L40

\bibitem[\protect\citeauthoryear{Boylan--Kolchin et al.}{2012}]{boylan2012}
Boylan--Kolchin, M., Bullock, J. S., \& Kaplinghat, M., 2012, MNRAS, 422, 1203

\bibitem[\protect\citeauthoryear{Burkert}{2000}]{burkert2000}
Burkert, A., 2000, ApJ, 534, L143

\bibitem[\protect\citeauthoryear{Cyr-Racine \& Sigurdson}{2013}]{cyrracine2013}
Cyr-Racine, F.-Y., \& Sigurdson, K., 2013, Phys. Rev. D, 87, 103515

\bibitem[\protect\citeauthoryear{Dav\'{e} et al.}{2001}]{dave2001}
Dav\'{e}, R., Spergel, D. N., Steinhardt, P. J., \& Wandelt, B. D., 2001, ApJ, 547, 574

\bibitem[\protect\citeauthoryear{de Blok et al.}{2001}]{deblok2001}
de Blok, W. J. G., McGaugh, S. S., Bosma, A., \& Rubin, V. C., 2001, ApJ, 552, L23

\bibitem[\protect\citeauthoryear{de Souza et al.}{2013}]{desouza2013}
de Souza, R. S., Mesinger, A., Ferrara, A., Haiman, Z., Perna, R., \& Yoshida, N., 2013, MNRAS, 432, 3218

\bibitem[\protect\citeauthoryear{de Vega et al.}{2013}]{devega2013}
de Vega, H. J., Falvella, M. C., \& Sanchez, N. G., 2013, arXiv:1307.1847

\bibitem[\protect\citeauthoryear{Dodelson \& Widrow}{1994}]{dodelson1994}
Dodelson, S., \& Widrow, L. M., 1994, Phys. Rev. Lett., 72, 17

\bibitem[\protect\citeauthoryear{Donato et al.}{2009}]{donato2009}
Donato, F., Gentile, G., Salucci, P., Frigerio Martins, C., Wilkinson, M. I., Gilmore, G., Grebel, E. K., Koch, A., \& Wyse, R., 2009, MNRAS, 397, 1169

\bibitem[\protect\citeauthoryear{Eisenstein \& Hu}{1998}]{eisenstein1998}
Eisenstein, D. J., \& Hu, W., 1998, ApJ, 496, 605

\bibitem[\protect\citeauthoryear{Field}{1958}]{field1958}
Field, G. B., 1958, Proc. I.R.E., 46, 240

\bibitem[\protect\citeauthoryear{Furlanetto et al.}{2006}]{furlanetto2006}
Furlanetto, S. R., Peng Oh, S., \& Briggs, F. H., 2006, Phys. Rep., 433, 181

\bibitem[\protect\citeauthoryear{Garrison-Kimmel et al.}{2013}]{kimmel2013}
Garrison-Kimmel, S., Rocha, M., Boylan-Kolchin, M., Bullock, J., \& Lally, J., 2013, arXiv:1301.3137

\bibitem[\protect\citeauthoryear{Governato et al.}{2007}]{governato2007}
Governato, F., Willman, B., Mayer, L., Brooks, A., Stinson, G., Valenzuela, O., Wadsley, J., \& Quinn, T., 2007, MNRAS 374, 1479

\bibitem[\protect\citeauthoryear{Haiman et al.}{2000}]{haiman2000}
Haiman, Z., Abel, T., \& Rees, M. J., 2000, ApJ, 534, 11

\bibitem[\protect\citeauthoryear{Hirata et al.}{2006}]{hirata2006}
Hirata C. M., 2006, MNRAS, 367, 259

\bibitem[\protect\citeauthoryear{Jenkins et al.}{2001}]{jenkins2001}
Jenkins, A., Frenk, C. S., White, S. D. M., Colberg, J. M., Cole, S., Evrard, A. E., Couchman, H. M. P., \& Yoshida, N., 2001,  MNRAS, 321, 372

\bibitem[\protect\citeauthoryear{Kaplan et al.}{2010}]{kaplan2010}
Kaplan, D. E., Krnjaic, G. Z., Rehermann, K. R., \& Wells, C. M., 2010, JCAP, 2010, 021

\bibitem[\protect\citeauthoryear{Kang et al.}{2013}]{kang2013}
Kang, X., Macci˜, A. V., \& Dutton, A. A., 2013, ApJ, 767, 22

\bibitem[\protect\citeauthoryear{Klypin et al.}{1999}]{klypin1999}
Klypin, A., Kravtsov, A. V., Valenzuela, O., \& Prada, F., 1999, ApJ, 522, 82

\bibitem[\protect\citeauthoryear{Lewis \& Challinor}{2007}]{lewis2007}
Lewis, A., \& Challinor, A., 2007, Phys. Rev. D, 76, 083005

\bibitem[\protect\citeauthoryear{Loeb \& Zaldarriaga}{2004}]{loeb2004}
Loeb, A., \& Zaldarriaga, M., 2004, Phys. Rev. Lett., 92, 211301

\bibitem[\protect\citeauthoryear{Lovell et al.}{2012}]{lovell2012}
Lovell, M. R., Eke, V., Frenk, C. S., Gao, L., Jenkins, A., Theuns, T., Wang, J., White, S. D. M., Boyarsky, A., \& Ruchayskiy, O., 2012, MNRAS, 420, 2318

\bibitem[\protect\citeauthoryear{Macci˜ et al.}{2012}]{maccio2012}
Macci\`{o}, A. V., Paduroiu, S., Anderhalden, D., Schneider, A., \& Moore, B., 2012, MNRAS, 424, 1105

\bibitem[\protect\citeauthoryear{Madau et al.}{1997}]{madau1997}
Madau, P., Meiksin, A., \& Rees, M. J., 1997, ApJ, 475, 429

\bibitem[\protect\citeauthoryear{Mapelli et al.}{2006}]{mapelli2006}
Mapelli, M., Ferrara, A., \& Pierpaoli, E., 2006, MNRAS, 369, 1719

\bibitem[\protect\citeauthoryear{Mesinger et al.}{2009}]{mesinger2009}
Mesinger, A., Bryan, G. L., \& Haiman, Z, 2009, MNRAS, 399, 1650

\bibitem[\protect\citeauthoryear{Mesinger et al.}{2013a}]{mesinger2013a}
Mesinger, A., Ewall-Wice, A., \& Hewitt, J., 2013a, arXiv:1310.0465

\bibitem[\protect\citeauthoryear{Mesinger et al.}{2013b}]{mesinger2013b}
Mesinger, A., Ferrara, A., \& Spiegel, D. S., 2013b, MNRAS, 431, 621

\bibitem[\protect\citeauthoryear{Mesinger et al.}{2011}]{mesinger2011}
Mesinger, A., Furlanetto, S., \& Cen, R., 2011, MNRAS, 411, 955

\bibitem[\protect\citeauthoryear{Mesinger et al.}{2005}]{mesinger2005}
Mesinger, A., Perna, R., \& Haiman, Z., 2005, ApJ, 623, 1

\bibitem[\protect\citeauthoryear{Moore et al.}{1999}]{moore1999}
Moore, B., Ghigna, S., Governato, F., Lake, G., Quinn, T., Stadel, J., \& Tozzi, P., 1999, ApJ, 524, L19

\bibitem[\protect\citeauthoryear{Morales \& Wyithe}{2010}]{morales2010}
Morales, M. F., \& Wyithe, J. S. B., 2010, ARA\&A, 48, 127

\bibitem[\protect\citeauthoryear{Naoz \& Barkana}{2005}]{naoz2005}
Naoz, S., \& Barkana, R., 2005, MNRAS, 362, 1047

\bibitem[\protect\citeauthoryear{Narayanan et al.}{2000}]{narayanan2000}
Narayanan, V. K., Spergel, D. N., DavŽ, R., \& Ma, C. P., 2000, ApJ, 543, L103

\bibitem[\protect\citeauthoryear{Newman et al.}{2009}]{newman2009}
Newman, A. B., Treu, T., Ellis, R. S., Sand, D. J., Richard, J., Marshall, P. J., Capak, P., \& Miyazaki, S., 2009, ApJ, 706, 1078

\bibitem[\protect\citeauthoryear{Pacucci et al.}{2013}]{pacucci2013}
Pacucci, F., Mesinger, A., \& Haiman, Z., 2013, arXiv:1306.0009

\bibitem[\protect\citeauthoryear{Pagels \& Primack}{1982}]{pagels1982}
Pagels, H., \& Primack, J. R., 1982, Phys. Rev. Lett., 48, 223

\bibitem[\protect\citeauthoryear{Papastergis et al.}{2011}]{Papastergis2011}
Papastergis E., Martin A. M., Giovanelli R., Haynes M. P., 2011, ApJ, 739, 38

\bibitem[\protect\citeauthoryear{Peebles}{2001}]{peebles2001}
Peebles, P. J. E., 2001, ApJ, 557, 495

\bibitem[\protect\citeauthoryear{Pober et al.}{2013}]{pober2013}
Pober, J. C., et al., 2013, ApJ, 768, L36

\bibitem[\protect\citeauthoryear{Pontzen \& Governato}{2012}]{pontzen2012}
Pontzen, A., \& Governato, F., 2012, MNRAS, 421, 3464

\bibitem[\protect\citeauthoryear{Pritchard \& Furlanetto}{2007}]{pritchard2007}
Pritchard, J. R., \& Furlanetto, S. R., 2007, MNRAS, 376, 1680

\bibitem[\protect\citeauthoryear{Seljak et al.}{2006}]{seljak2006}
Seljak, U., Makarov, A., McDonald, P., \& Trac, H., 2006, Phys. Rev. Lett., 97, 191303

\bibitem[\protect\citeauthoryear{Sheth et al.}{2001}]{sheth2001}
Sheth, R. K., Mo, H. J., \& Tormen, G., 2001, MNRAS, 323, 1

\bibitem[\protect\citeauthoryear{Smith \& Markovic}{2011}]{smith2011}
Smith, R. E., \& Markovic, K., 2011, Phys. Rev. D, 84, 063507

\bibitem[\protect\citeauthoryear{Sobacchi et al.}{2013}]{sobacchi2013}
Sobacchi, E., \& Mesinger, A., 2013, MNRAS, 432, 3340

\bibitem[\protect\citeauthoryear{Spergel \& Steinhardt}{2000}]{spergel2000}
Spergel, D. N., \& Steinhardt, P. J., 2000, Phys. Rev. Lett., 84, 3760

\bibitem[\protect\citeauthoryear{Teyssier et al.}{2013}]{teyssier2013}
Teyssier, R., Pontzen, A., Dubois, Y., \& Read, J. I., 2013, MNRAS, 429, 3068

\bibitem[\protect\citeauthoryear{Vald\'{e}s et al.}{2013}]{valdes2013}
Vald\'{e}s, M., Evoli, C., Mesinger, A., Ferrara, A., \& Yoshida, N., 2013, MNRAS, 429, 1705

\bibitem[\protect\citeauthoryear{Viel et al.}{2008}]{viel2008}
Viel, M., Becker, G. D., Bolton, J. S., Haehnelt, M. G., Rauch, M., \& Sargent, W. L., 2008, Phys. Rev. Lett., 100, 041304

\bibitem[\protect\citeauthoryear{Viel et al.}{2005}]{viel2005}
Viel, M., Lesgourgues, J., Haehnelt, M. G., Matarrese, S., \& Riotto, A., 2005, Phys. Rev. D, 71, 063534

\bibitem[\protect\citeauthoryear{Viel et al.}{2013}]{viel2013}
Viel, M., Becker, G. D., Bolton, J. S., Haehnelt, M. G., 2013, arXiv:1306.2314

\bibitem[\protect\citeauthoryear{Villaescusa-Navarro \& Dalal}{2011}]{navarro2011}
Villaescusa-Navarro, F., \& Dalal, N., 2011, JCAP, 2011, 024

\bibitem[\protect\citeauthoryear{Wouthuysen}{1952}]{wouthuysen1952}
Wouthuysen, S. A., 1952, Astron. J., 57, 31

\bibitem[\protect\citeauthoryear{Zaldarriaga et al.}{2004}]{zaldarriaga2004}
Zaldarriaga, M., Furlanetto, S. R., \& Hernquist, L., 2004, ApJ, 608, 622

\end{thebibliography}
\end{document}